\documentclass[aps,floatfix,nofootinbib,preprint]{revtex4}

\usepackage{graphicx}
\usepackage{epic}
\usepackage{eepic}
\usepackage{latexsym}
\usepackage{amssymb,amsmath}

\newcommand{\eq}[1]{(\ref{#1})}
\newcommand{\be}{\begin{equation}}
\newcommand{\ee}{\end{equation}}
\newcommand{\bea}{\begin{eqnarray}}
\newcommand{\eea}{\end{eqnarray}}

\newcommand{\vs}[1]{\vspace{#1 mm}}
\newcommand{\hs}[1]{\hspace{#1 mm}}

\newcommand{\vc}{\vec{k}}
\newcommand{\vx}{\vec{x}}
\newcommand{\vv}{|0\right>}
\newcommand{\vvc}{\left<0|}

\newcommand{\oo}{\omega_0}

\def\h{\hat}

\def\cc{\gamma}

\def\d{\delta}

\def\e{\epsilon}

\def\f{\phi}
\def\fr{\frac}
\def\F{\Phi}
\def\vf{\varphi}
\def\h{\eta}

\def\l{\lambda}

\def\m{\mu}
\def\n{\nu}

\def\r{\rho}

\def\z{\zeta}

\def\o{\omega}

\def\del{\partial}

\let\bm=\bibitem
\def\nn{\nonumber}

\begin{document}

\title{Vacuum Polarization, Geodesic Equation and Sachs-Wolfe Effect} 

\author{Ali Kaya}

\email[]{alikaya@tamu.edu}
\affiliation{\vs{3}Department of Physics and Astronomy, Texas A\&M University, College Station, TX 77843, USA \vs{10}}

\begin{abstract}

We show that the null geodesic equation for photons is modified in the presence of a charged scalar field, with quantum fluctuations acting as an effective mass term that changes the null paths to timelike curves. This effect can be interpreted as a vacuum polarization phenomenon in curved spacetime. The resulting contribution to the Sachs-Wolfe effect varies with photon frequency, leading to frequency-dependent corrections to the cosmic microwave background (CMB) blackbody spectrum in the form of a $\mu$-distortion, as well as modifications to the CMB power spectrum.  We estimate these within a standard inflationary scenario and find that while the correction to the CMB power spectrum is significant when the scalar field is light,  the magnitude of the $\mu$-distortion depends strongly on the regularization prescription. 

\vs{5}
\end{abstract}

\maketitle

\section{Introduction} 

The quantum vacuum exhibits remarkable properties.  It effectively behaves as a {\it medium} due to the creation of virtual charged particles, a phenomenon known as vacuum polarization. In flat spacetime,  vacuum polarization gives rise to several observable effects including charge screening, running of the fine-structure constant, Lamb shift,  light-light scattering and spectral line shifts in exotic atoms. In an expanding universe, vacuum polarization effects have also been studied to investigate whether  the large scale micro-Gauss magnetic fields observed in galaxies and galaxy clusters could  be generated by vacuum fluctuations during inflation \cite{vp1,vp2}. 
 
In this work, we investigate the impact of vacuum polarization on the motion of photons within the geometric optics approximation. Since the vacuum effectively behaves like a medium, it is expected that the geodesic equation will be modified. Our goal is to determine whether this effect leaves an observable imprint on CMB photons as they travel from the last scattering surface to us.
 
 \section{Photon Propagation in a Scalar Field Medium} 
 
Let us first review how the geodesic equation follows from the field equations in the geometric optics approximation. While this is typically discussed for massless fields, we aim to generalize the result to massive particles (the following discussion for does not offer a genuine geometrical optics derivation of Maxwell’s equations, but it captures the main idea). The construction is most clearly illustrated for a minimally coupled massive test scalar field on a curved background obeying  
\be\label{kg} 
\nabla_\m \nabla^\m\vf-m^2\vf =0,
\ee
where $\nabla_\m$ is the (unique torsion free and metric compatible) covariant derivative of  $g_{\m\n}$ with signature $(-,+,+,+)$. To obtain the geodesic equation, one takes a solution in the form 
\be \label{vfs} 
\vf={\cal A} \exp(iS)
\ee
and supposes that the phase is highly oscillating. This amounts to assuming  that $|\nabla_\m S|$ is much larger than any other  relevant scale, such as the spacetime curvature. One can then setup a perturbation theory based on the number of $\nabla_\m S$ factors. Defining 
\be\label{ks} 
k_\mu\equiv \nabla_\m S
\ee
the leading order term in \eq{kg} implies 
\be\label{p1}
k^\m k_\m=-m^2,
\ee
which shows $k^\m$ is a time-like vector. Using $\nabla_\n(k^\m k_\m)=0$,  one can also see that $k^\m$ is a geodesic tangent $k^\m \nabla_\m k^\n=0$. Therefore, in this approximation, the surfaces of constant phase are generated by time-like geodesics. The next order equation dictates how the amplitude ${\cal A}$ changes along the geodesics 
\be\label{p2} 
k^\m \nabla_\m \ln {\cal A}=-\fr12 \nabla_\m k^\m.
\ee
To obtain a  perturbative series to all orders, one  introduces a formal expansion parameter $\e$  so that  $\vf=({\cal A}+\e {\cal B} + {\cal O} (\e^2)) \exp(iS/\e)$ \cite{mtw}. Eq. \eq{p1} and \eq{p2} follow from orders $1/\e^2$ and $1/\e$, respectively, and ${\cal B}$ is determined from $\e^0$ equation, etc. 

There is a crucial subtlety in the above derivation; i.e. $m^2$ is assumed to be of the highest order $1/\e^2$, which is not immediately obvious. In \cite{geroch}, it was shown that small bodies in general relativity follow null or timelike geodesics. This result applies to wave packets of a massive Klein-Gordon field only when $m$ increases, as this physically enhances the formation of wave packets. Technically, in the proof presented in  \cite{geroch} only the collection of Klein-Gordon solutions for all mass parameter, rather than for a fixed value, tracks timelike geodesics, see \cite{jow} for a brief summary.  These findings are consistent with the assumption $m^2= {\cal O}(1/\e^2)$. Moreover, in the flat-space limit, where curvature is turned off in accordance with the geometric optics approximation, \eqref{p1} is the standard relation that one would like to obtain. 

Let us now consider the following Lagrangian of a charged (complex) scalar field and the electromagnetic field in a curved spacetime 
\be
L=-\fr14\sqrt{-g}F^{\m\n}F_{\m\n}-\sqrt{-g}(D_\m\f) (D^\m\f)^*-m^2\sqrt{-g}\f^*\f,\label{1}
\ee
where $F_{\m\n}=\del_\m A_\n-\del_\n A_\m$, $g=\det(g_{\m\n})$, $D_\m\f=\del_\m\f-ieA_\m\f$ is the covariant derivative and $e$ is the dimensionless scalar-photon coupling constant. The field equations imply 
\be\label{fmn}
\nabla^\m F_{\m\n}=j_\n,
\ee
where
\be\label{jn} 
j_\n=-i e \left(\f\del_\n \f^*-\f^*\del_\n \f\right)+2e^2(\f^*\f)A_\n. 
\ee
Note that $j_\n$ is real. 

We are interested in the lowest order impact of  the scalar field (vacuum) fluctuations on the motion of photons created, say during reheating after inflation. The first term in \eq{jn} is a non-homogeneous source term that produces a particular field $A_{P}^\m=-ie\int G^{\m\n}(\f\del_\n \f^*-\f^*\del_\n \f)$, where $G^{\m\n}$ is a Green function in an appropriate gauge obeying certain boundary conditions.  At this linear order, $A^\m_P$ does not interact with the photons generated by other sources.

Obviously, the  second term in \eq{jn} appears like an effective mass term in \eq{fmn}, which should modify photon propagation. To determine its effect in the geometric optics approximation, one can proceed as in the case of  massive Klein-Gordon field discussed above. After introducing the phase and the amplitude as
\be
A_\m=a_\m \exp(iS),
\ee
the field equations can be analyzed order by order by assuming the phase is highly oscillating. Imposing the gauge $\nabla^\m A_\m=0$, which implies at the leading order $k^\m a_\m=0$, one can obtain\footnote{Note that the field equations \eq{fmn} yield a curvature term $R_{\m\n}a^\n$ arising from the commutator of the covariant derivatives, where $R_{\m\n}$ is the Ricci tensor. Although this can also be viewed as a mass term, we assume all such curvature corrections are small and only modify higher order equations. This is appropriate since by Einstein's equations the curvature is suppressed by the Planck mass as compared to energy density of matter.} 
\bea
&&k^\m k_\m=-2e^2\f^*\f,\nn\\
&& k^\m\nabla_\m k_\n=-e^2\nabla_\n(\f^*\f), \label{mg}
\eea
where, as in \eq{ks}, $k_\mu= \nabla_\m S $. 

Eq. \eq{mg} shows that photons now follow {\it non-geodesic timelike curves,} behaving as if they propagate through a medium. Note that the two equations in \eq{mg} are consistent; indeed the first one can be seen to hold through a path if it is satisfied at an initial point. In deriving \eq{mg} from the field equations, one uses  $k_\mu= \nabla_\m S $, i.e. $k^\m$ is hypersurface orthogonal.  In solving \eq{mg} below we will ignore this fact. Indeed, one can show that when $(\nabla_\m k_\n-\nabla_\n k_\m)=0$ at a point $P$, then $k^\r\nabla_\r(\nabla_\m k_\n-\nabla_\n k_\m)(P)=0$. As a result, if a congruence of curves defined by \eqref{mg} is initially hypersurface orthogonal, it will remain so, at least within an initial neighborhood. The same holds true when the right-hand side of \eqref{mg} is zero.

\begin{figure}
	\centerline{\includegraphics[width=7cm]{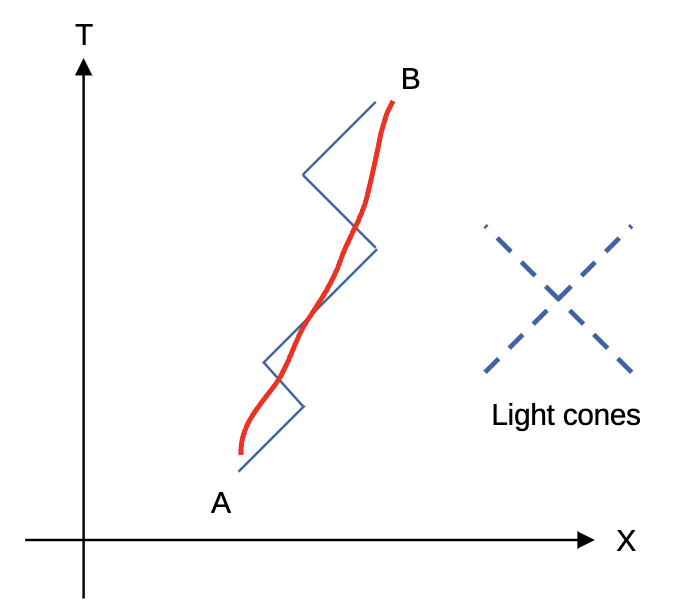}}
	\caption{An illustration of photon propagation from point A to point B in two-dimensional spacetime.  The actual path is (piece-wise) null (between collisions) while the mean path becomes timelike.} 
	\label{fig1}
\end{figure}

It is crucial to emphasize that the above analysis does not imply  photons acquire mass at the fundamental level. Eq. \eq{mg} is an effective mean description of photon trajectories. This is similar to light propagation in a medium where the photons does not acquire a mass but still  their speed in the medium is less than the speed of light in vacuum. In such cases, photons scatter off charged particles making their average path timelike, see FIG \ref{fig1}. 

We will determine solutions of \eq{mg} in  the following perturbed cosmological spacetime 
\be\label{met}
ds^2=a(\h)^2\left[-(1+2\Psi)d\h^2+[(1+2\Phi)\d_{ij}+\cc_{ij}]dx^i dx^j\right],
\ee
by treating the charged scalar field contribution as a small correction. The tensor mode is traceless $\d^{ij}\cc_{ij}=0$ but there is  no need to  impose any gauge fixing conditions. For generality, the scalar modes $\Psi$ and $\Phi$ are taken to be different from each other. 

In \cite{akg1,akg2}, we have studied the solutions of the null geodesics in the background \eq{met}. Here, we generalize these findings for \eq{mg}.  Because the impact of the charged scalar fluctuations on the photon propagation is assumed to be small, we consider a curve perturbed around a null geodesic of the unperturbed spacetime. The tangent vector $k^\m$ of such a curve can be written as 
\bea
&&k^0=\fr{\oo}{a^2}+\d k^0,\nn \\
&&k^i=\oo\fr{ l^i}{a^2}+\d k^i,\hs{5}\d_{ij}l^i l^j=l^il^i=1,\label{g} 
\eea
where $(0,i)$ are coordinate indices referring to $(\h,x^i)$, and  $\oo$ and $l^i$ are constants. Note that $\oo/a$ is the frequency observed by a comoving observer in the unperturbed spacetime and $l^i$ defines the spatial direction of propagation.  

To proceed, one can determine $\d k^0$ from the first equation in \eq{mg} so that 
\be\label{dk0}
\d k^0=l^i\d k^i+\fr{\oo}{a^2}\left(\Phi-\Psi\right)+\fr{\oo}{2a^2}\cc_{ij}l^i l^j+\fr{1}{\oo}e^2\f^*\f. 
\ee
Then, $\d k^i$ can be solved from the second equation in \eq{mg} as  
\be
\d k^i(x,\h)=\d k^i_{int}+\d k^i_{H}
-\fr{2\oo}{a(\h)^2}l^i\,\Phi(x,\h)-\fr{\oo}{a(\h)^2}\cc_{ij}(x,\h)\,l^j . \label{dki} 
\ee
Here, $\d k^i_{int}$ is the piece of the solution that can be expressed as an integral over the unperturbed null geodesic path so that
\be
\d k^i_{int}(x^i,\h)=\fr{\oo}{a(\h)^2}\del_i\left\{\int_{\h_0}^{\h} d\h'\left[\Phi- \Psi+\fr12 \cc_{jk}l^j l^k
-\fr{e^2a(\h')^2}{\oo^2}(\phi^*\phi)\right]\left(x^i_{\h'\h},\h'\right)\right\},\label{dkint}
\ee
where $\h_0$ is the present conformal time and $x^i_{\h_1\h_2}$ is the spatial position on the unperturbed null geodesic 
\be\label{xl}
x_{\h_1\h_2}^i=x^i+l^i(\h_1-\h_2).
\ee
In \eq{dkint}, $\left(x^i_{\h'\h},\h'\right)$ is the argument of the fields inside the square bracket, which must be integrated with respect to $\h'$. 

On the other hand, $\d k^i_{H}=F^i(x^j-\h l^j)/a^2$ with $F^i$ arbitrary is the homogeneous solution obeying
\be
\left(\del_\h+l^j\del_j\right)\d k^i_{H}=0. 
\ee
By choosing 
\be
\d k^i_{H}(x^i,\h)=\fr{\oo}{a(\h)^2}\left[2l^i\,\Phi+\cc_{ij}\,l^j\right](x^i_{\h_0\h},\h_0)\label{kh}
\ee
one can obtain the unique solution with 
\be\label{ic0}
\d k^i(x,\h_0)=0,
\ee
so that $l^i$ defines the present direction of propagation, i.e.  the actual line of sight. 

\section{Sachs-Wolfe Effect} 

The equations \eqref{g}, \eqref{dk0}, and \eqref{dki} describe the perturbed photon trajectory in the spacetime \eqref{met} under the influence of the charged scalar field $\f$ and the metric perturbations. It is important  to determine the frequency of these photons as measured by a  comoving observer in \eq{met}.  The 4-velocity vector $u^\m$ of such an observer is given by
\be\label{u} 
u^0=\fr{1}{a}(1-\Psi),\hs{5}u^i=0.
\ee
Note that $u^\m u_\m=-1$ to the lowest order in perturbations. The frequency measured by this observer equals
\be\label{up}
\o=-u^\m k_\m,
\ee
which can be calculated as
\be
\o=\fr{\oo}{a}\left[1+\fr{a^2 }{\oo}l^i(\d k^i_{int}+\d k^i_{H})-\fr12\cc_{ij}l^i l^j-\F
+e^2\fr{a^2}{\oo^2}\f^*\f\,\right].\label{w} 
\ee
One can verify that all the terms inside the parentheses above are dimensionless, as expected.

It was shown in \cite{rd,akg1} that \eq{w} directly implies the Sachs-Wolfe effect. To see this in our case, we first note that \eq{mg} can be derived from the action
\be\label{slxk}
S=\int d\tau\left[k_\m\fr{dx^\m}{d\tau}-\l\left(\fr12 g^{\m\n}\,k_\m k_\n-e^2\f^*\f\right)\right],
\ee
where $\l$ is a Lagrange multiplier, and  $(x^\m,k_\n)$ are conjugate canonical variables. Therefore,  Liouville's theorem is applicable in the phase space. The invariance of \eq{slxk} under  $\tau$-reparametrization can be fixed by imposing the gauge $x^0=\h=\tau$ and solving $k_0$ from the constraint. This deparametrizes the system leaving $(x^i,k_j)$ as dynamical variables and $\h$ as the time coordinate. In this case, Liouville's theorem applies to phase space distribution functions of the form $f(x^i,k_j,\h)$. Now, for a distribution function given by $f=f(\o/T)$, where $\o=g_{ij}k^i k^j$,  Liouville's theorem imposes 
\be\label{tw}
\fr{\d T}{T}=\fr{\d \o}{\o}. 
\ee
One can see that $\o$, as given in \eq{w}, can be written as  $\o=g_{ij}k^i k^j$ to first order in perturbations (recall that $\o$ was originally defined by \eq{up} as the frequency observed by a comoving observer). From \eq{w} and \eq{tw}, one can then find  
\be\label{sw}
\left(\del_\h+l^j\del_j\right)\fr{\d T}{T}=-\Phi'-l^i\del_i\Psi-\fr12 \cc_{ij}'l^il^j + \fr{e^2}{\oo^2}\left(a^2\f^*\f\right)',
\ee
where a prime denotes an $\h$-derivative. The first three terms in the right hand side of \eq{sw} give the standard Sachs-Wolfe effect (see e.g. \cite{mukhanov}), and the last term is the contribution of the charged scalar field. 

It is peculiar that the last term in \eqref{sw} depends on $\oo$.  In the standard cosmological model, CMB temperature fluctuations are frequency-independent, with any observed frequency dependence typically attributed to foreground contamination. However, there are known mechanisms that can introduce frequency dependence. For instance,  CMB spectral distortions (i.e. small deviations from the perfect black-body spectrum) can arise from effects such as annihilating dark matter or decaying relic particles, which can induce frequency dependence, see \cite{cd1,cd2,cd3} for review. Similarly, primordial magnetic fields can also generate  frequency-dependent distortions or anisotropies in the CMB, see \cite{fd1} for a review. Clearly, these mechanisms differ significantly from the one discussed here.

\section{$\m$-Distortion}

We now use the above results to track the evolution of the CMB spectrum from the last scattering surface to the present day.  Let us first consider the one-point function $\left<\d T/T\right>$. Any frequency-independent change in \eqref{w} preserves the blackbody shape, as is the case for the standard Sachs-Wolfe terms involving gravitational perturbations (moreover, these vanish in the one-point function since e.g. $\left<\Phi\right>=0$ for quantum fluctuations). On the contrary, the last term in \eq{w} distorts the blackbody shape. 

To determine the distortion, one can use \eq{w} to express the photon frequencies at the last scattering surface and today, denoted by $\o_{L}$ and $\o_{obs}$,  respectively. This gives 
\bea
&&\o_{obs}=\o_0\left[1+\fr{e^2}{\oo^2}\f^*\f(\h_0)\right]\nn\\
&&\o_L=\fr{\o_0}{a_L}\left[1+e^2\fr{a_L^2}{\oo^2}\f^*\f(\h_L)\right]\label{wobsl}
\eea
where $a_L$ is the scale factor at the last scattering and we normalize $a(\h_0)=1$. In \eq{wobsl}, only the dominant scalar-dependent terms are shown. An integrated scalar field term originates from \eq{dkint}, but it is presumably suppressed, \footnote{\label{f2}It may be worth noting that this contribution may not be  entirely negligible as in the case with the  Integrated-Sachs-Wolfe effect, which  contributes a detectable fraction to Sachs-Wolfe effect,.} while the remaining contributions are the well-known gravitational ones. Eq.  \eq{tw} implies
\be
\fr{\o_L}{T_L}=\fr{\o_{obs}}{T},
\ee
where $T_L$ is the temperature at the last scattering surface. This yields to the first order in corrections  
\be 
T=T_0\left[1+\fr{e^2}{\oo^2}\f^*\f(\h_0)-e^2\fr{a_L^2}{\oo^2}\f^*\f(\h_L)\right],
\ee
where $T_0=T_L a_L$ is the current  CMB temperature. Defining the $\m$-distortion in the Boltzmann factor as 
\be
\exp(\o_{obs}/T)=\exp(\o_{obs}/T_0+\m)
\ee 
one can find (note that to the lowest order $\o_{obs}=\oo$)
\be\label{mmon}
\m=\fr{e^2}{\oo\,T_0}\left[a_L^2\left<\f^*\f\right>_L-\left<\f^*\f\right>_0\right],
\ee
where we replace the field values with $\left<\f^*\f\right>_L$ and $\left<\f^*\f\right>_0$;  the vacuum expectation values at the last scattering surface and today, respectively.

A key issue is now to determine $\left<\f^*\f\right>_L$ and $\left<\f^*\f\right>_0$. We estimate these quantities within the free theory—neglecting any loop contributions from interactions—by assuming that the scalar field was initially in its Bunch-Davies vacuum during inflation in the early universe. Naturally, these expectation values exhibit UV divergences since the field operators have coincident spacetime coordinates and thus they must be regularized. 

A standard method is to use dimensional regularization; however, it is difficult to apply in this context, as the analytical form of the Bunch-Davies mode functions {\it after inflation} is challenging to determine (the mode functions during a de Sitter phase are given by Hankel functions). Moreover, dimensional regularization requires that renormalization conditions be imposed to fix counter-terms, which typically give rise to logarithmic dependence on the renormalization scale (see, e.g., the appendix of \cite{dr-kaya} for a discussion of how dimensional regularization works for the two-point function in de Sitter space). Yet, the Bunch-Davies mode functions exhibit the same UV behavior as their flat-space counterparts, and on dimensional grounds, one therefore expects $\left<\f^*\f\right>\sim m^2$.  

An alternative approach is the adiabatic regularization \cite{ad1} as applied to the two point function \cite{ad2}, where  one subtracts the adiabatic mode contributions from the Bunch-Davies modes (note that this method is more suited to cosmological settings and does not require any renormalization conditions to be imposed). The evolution of Bunch-Davies mode function after inflation depends on its comoving wave-number $k$. For instance,  modes that satisfy $k/a > \textrm{max}(m,H)$ at all times evolve adiabatically, and their contribution to the two-point function is canceled by adiabatic subtractions. One can then see that, since $m\gg H$ during the period of interest, adiabatic regularization also yields $\left<\f^*\f\right>\sim m^2$.

As a result  for $m\sim$ TeV, which is a reasonably small mass scale, one finds $\m\gg1$ unless $e$ is extremely small (the current bound on $\m$ is $|\m|<9\times 10^{-5}$ from COBE/FIRAS and proposed experiments like PIXIE aim to to push this bound to $|\m|\sim10^{-8}$). 

A large $\m$ appears from the same root cause as the cosmological constant problem, i.e. there is a fundamental lack of our understanding about the cosmological impact of coincident vacuum expectation values. One can try to estimate $\left<\f^*\f\right>$  by applying the same amount of mismatch encountered in the cosmological constant problem. The naive dimensional or adiabatic regularization gives $\left<\rho\right>\sim m^4$, where $\r$ is the energy density, and assuming this must have  the same order of magnitude as the dark energy today gives the rough mismatch factor $H_0^2M_p^2/m^4$, where $H_0$ is the current Hubble value. Because $\r$ has mass dimension 4 and $\f^*\f$ has mass dimension 2, the mismatch for $\left<\f^*\f\right>_0$  can be $H_0M_p/m^2$, which implies $\left<\f^*\f\right>_0\sim H_0M_p$. Taking $\oo\sim T_0\sim 10^{-4} eV$, the second term in \eq{mmon} gives $\m\sim 10^{-5} e^2$, which can be within the reach of the upcoming measurements if $e\sim 1/10$. 

\section{Power Spectrum} 

Let us now consider the two-point angular correlation function
\be\label{tt} 
\left<\fr{\d T(\hat{l}_1)}{T_0}\fr{\d T(\hat{l}_2)}{T_0}\right>,
\ee
where $\hat{l}_1$ and $\hat{l}_2$ denote two different directions in space. One can use \eq{w} and \eq{tw} to relate \eq{tt} to the correlation functions of the fields $\F$, $\Psi$, $\cc_{ij}$ and $\f$. The contribution of $\d k^i_{H}$ in \eq{w} is completely negligible since the fields are evaluated today (the time argument of the perturbations in \eq{kh} is  $\h_0$). On the other hand, by \eq{dkint}, $\d k_{int}$ involves a spatial derivative of an integral extending from the last scattering surface to us, which is expected to be small since the fluctuations do not change appreciably over this range (see, however, footnote \ref{f2}). Therefore, the largest contribution to \eq{tt} comes from $\F$ and $e^2 a^2 \f^*\f/\oo^2$ terms in \eq{w} 

To compare the contributions of these terms, one can calculate the equal time correlation functions $\left<\Phi(x_1)\Phi(x_2)\right>$ and $\left<\f^*\f(x_1)\f^*\f(x_2)\right>$, where $x_1$ and $x_2$ denote the spatial coordinates of two points on the last scattering surface projected from the directions $\hat{l}_1$ and $\hat{l}_2$, respectively (care must be taken when projecting a line of sight to the last scattering surface due to gravitational lensing, see  \cite{lensing} for a review). As it is standard in calculating cosmologically relevant correlation functions, one should account for the contribution of modes whose (comoving) wavelengths satisfy $\l>|x_1-x_2|$, assuming that shorter wavelengths average out to zero due to oscillations. 

We determine $\left<\Phi(x_1)\Phi(x_2)\right>$ and $\left<\f^*\f(x_1)\f^*\f(x_2)\right>$ at superhorizon scales in a  typical inflationary model, where $m$ is assumed to be much smaller than $H_I$, the Hubble scale of inflation, so that cosmologically relevant $\f$ modes can be generated. We imagine $\f$ is not the inflaton but rather it behaves like a test scalar during inflation. It is  convenient to compare the correlation functions in position space as they may remain well-defined in position space even when becoming singular in momentum space, see e.g. \cite{etc}. 

Since  $m\ll H_I$,  $\f$-modes that become superhorizon during inflation  have a scale free spectrum (assuming they have been released in the Bunch-Davies vacuum when they were subhorizon). A {\it real} scalar field operator $\vf$  can be Fourier decomposed as 
\be
\vf=\fr{1}{(2\pi)^{3/2}}\int d^3k\left[e^{i\vc.\vx}\,\vf_k(\h)\,a_{\vc}+h.c\right],\label{vfmodes}
\ee
where $h.c.$ stands for Hermitian conjugate, $\vf_k$ are mode functions and the vacuum is defined by $\left.a_{\vc}\vv=0$. The complex field $\f$ can be written as the sum of two real fields $\f=(\vf_1+i\vf_2)/\sqrt{2}$ (so that $\vf_{1,2}$ are canonically normalized in \eq{1}). Each of the fields $\vf_1$ and $\vf_2$ can be quantized as in \eq{vfmodes} with their own ladder operators $a^{(1)}_{\vc}$ and  $a^{(2)}_{\vc}$ (yet their mode functions are identical to $\vf_k$).  Using Wick's theorem one can see
\be
\vvc\f^*(x_1)\f(x_1)\f^*(x_2)\f(x_2)\vv=
(\vvc\vf(x_1)\vf(x_2)\vv)^2+\vvc\vf^2\vv^2,\label{ffnc}
\ee
where $\vvc\vf^2\vv=\vvc\vf(x)^2\vv$. Note that 
\be
\vvc\vf_{1,2}(x_{1,2})^2\vv=\vvc\vf(x)^2\vv=\fr{1}{(2\pi)^3}\int d^3 k |\vf_k|^2. 
\ee
We are going to ignore the last term in \eq{ffnc} since it is independent of  the spatial coordinates $\vx_1$ and $\vx_2$ (and depends only on $\h$), consequently it only changes the normalization of the spectrum (similarly, $\d T(\hat{l}_1)$ also involves today's vacuum expectation value $\vvc\f^*\f\vv_0$, which we disregard).  

Taking the standard scale free spectrum, which can be obtained from the Bunch-Davies mode function during inflation, one can write 
\be\label{vfk} 
\vf_k=Z\fr{H_I}{\sqrt{2k^3}},
\ee
where $Z$ is introduced to  take into account the possible change of the amplitude as the superhorizon mode evolves from the end of inflation to the last scattering surface ($Z=1$ if $\vf$ is massless). Using \eq{vfmodes}, \eq{vfk} and $a_{\vec{k}}\left. \vv=0$, one finds
\be\label{vfvf}
\vvc\vf(x_1)\vf(x_2)\vv=Z^2\fr{H_I^2}{4\pi^2}C,
\ee
where $C$ is a dimensionless number given  by
\be
C=\int_0^\infty d{\cal K}\,\fr{\sin({\cal K})}{{\cal K}^2}.
\ee
This integral is logarithmically divergent as ${\cal K}\to0$, which  is the usual infrared (IR) divergence that can be cured either by an  IR cutoff or by a tilt in the spectrum. In any case, $C$ can be viewed as an order unity constant due to the logarithmic behavior. Because $\vvc\vf(x_1)\vf(x_2)\vv$ does not depend on $|x_1-x_2|$ when  $x_1$ and $x_2$ are separated at {\it superhorizon distances}\footnote{Strictly speaking, \eq{vfvf} is valid at superhorizon separations as \eq{vfk} represents the superhorizon spectrum. In general, after performing angular integrations, the two-point function can be written as $\vvc\vf(x_1)\vf(x_2)\vv=1/2\pi^2\int_0^\infty dk k^2   |\vf_k|^2 \sin(kr)/(kr)$, where $r=|\vx_1-\vx_2 |$. This integral becomes negligible for $kr\gg1$ because of the oscillating behavior of the sine function, hence for a given $r$, only modes with $k\leq 1/r$ significantly contributes. When $r$ is superhorizon, one can thus use the superhorizon spectrum \eq{vfk} to evaluate  the two-point function, which leads to \eq{vfvf}.} from \eq{vfvf}, $\vvc\f^*(x_1)\f(x_1)\f^*(x_2)\f(x_2)\vv$ also has this property by \eq{ffnc}, therefore the contribution of $\f^*\f$ in \eq{w} to the CMB power spectrum is scale free at superhorizon distances. 

To determine  $Z$ in \eq{vfk} we examine the mode equation 
\be\label{me} 
\ddot{\vf}_k+3H\dot{\vf}_k+\fr{k^2}{a^2}\vf_k+m^2\vf_k=0,
\ee
where dot denotes derivative with respect to standard cosmic time  $dt=ad\h$. Since we are interested in superhorizon modes, the third term \eq{me} is negligible. In epochs when $m<H$, the mass term is also negligible and $\vf_k$ is approximately constant giving $Z\simeq1$. When $m>H$, the mode function starts to oscillate with the amplitude slowly decreasing. The solution can be written as $\vf_k\simeq A \sin(mt)$, where $A$ obeys $\dot{A}\simeq-3/2 H A$, which can be solved as $A\simeq a^{-3/2}$. As a result one finds for the overall amplitude 
\be\label{ae} 
A\simeq \begin{cases} 1,\hs{10}m<H\\ a^{-3/2},\hs{4} m>H.\end{cases}
\ee
The fact that superhorizon $\vf_k$ modes coherently oscillates does not diminish their contribution to Sachs-Wolfe effect since  the correction depends on the field square $\f^*\f$ and one observes many photons at different times which would yield a time average of, e.g.  $\sin^2(mt)$. Assuming for simplicity that when $m>H$ the universe has been matter dominated $a=t^{2/3}$, \eq{ae} gives 
\be\label{ahm} 
Z=\fr{H_L}{m},
\ee
where $H_L$ is the Hubble scale at the time of last scattering. The equations \eq{ahm} and  \eq{vfvf} can be used in \eq{ffnc} to calculate $\vvc\f^*(x_1)\f(x_1)\f^*(x_2)\f(x_2)\vv$. 

The main gravitational contribution to  Sachs-Wolfe effect, which we aim to compare with the vacuum polarization correction, arises from the correlation function $\vvc\F(x_1)\F(x_2)\vv$. The mode function $\F_k$ of $\F$ is related to the curvature perturbation\footnote{ Following \cite{mal}, we define the curvature perturbation from the Arnowitt–Deser–Misner (ADM) decomposition  $ds^2=-N^2dt^2+a^2\exp(2\z)\cc^{adm}_{ij}(dx^i+N^idt)(dx^j+N^jdt)$ where  $\det(\cc^{adm}_{ij}=1$). At the linear order $\cc^{adm}_{ij}=\d_{ij}+\cc_{ij}$, where $\cc_{ij}$ is introduced in \eq{met}.}
 mode function $\z_k$ during matter domination after inflation as $\F_k=3/5\,\z_k$. The mode function $\z_k$ is given by 
\be
\z_k=\fr{1}{M_p\sqrt{2\e}}\fr{H_I}{\sqrt{2k^3}},
\ee
where $\e$ is the slow-roll parameter and $M_p$ is the Planck mass. Using these one can calculate the well-known two point function (see e.g. \cite{mukhanov}) at superhorizon scales as 
\be\label{ff35}
\vvc\F(x_1)\F(x_2)\vv=\fr{9}{25}\fr{1}{2\e M_p^2}\fr{H_I^2}{4\pi^2}C.
\ee
Note that there is no amplitude correction $Z$ in \eq{ff35} as compared to \eq{vfvf}. 

It is now possible to compare the contributions of $\F$ and $e^2a^2\f^*\f/\oo^2$ in \eq{w} to the CMB power spectrum at superhorizon scales from their two point functions using \eq{ffnc}, \eq{vfvf} and \eq{ff35}. One finds they become equal when  
\be\label{cf}
\fr{\oo}{a_L}=\left(\fr{5}{3\pi\sqrt{2C}}\right)^{1/2}e\e^{1/4}\fr{H_L}{m}\sqrt{H_I M_p},
\ee
and the vacuum polarization contribution becomes larger if the photon frequency is smaller than \eq{cf}. From \eq{ae}, $H_L/m$ in \eq{cf} should be set to 1 if $m<H_L$. As noted before,  $\oo$ can be identified as the present frequency of photons and $\oo/a_L$ becomes their frequency at the last scattering surface. 

We know that  $a_L\simeq 10^{-3}$ and $H_L\simeq 10^{-28}$ eV. Assuming  $e\sim 1/10$, $\e\sim 10^{-4}$, $H_I\sim 10^{14}$ GeV and $m\sim 10^{-4}$ eV, one finds $\oo\sim 10^{-4}$ eV, which corresponds to the typical frequency of observed CMB photons. Thus, for $m\gg 10^{-4}$ eV, one has $\oo\ll  10^{-4}$ eV. In particular for $m\sim$ TeV, \eq{cf} falls several orders of magnitude below the frequency range detectable by current CMB telescopes.

\section{Conclusions} 

In a way, the above result  is not surprising and aligns with the standard lore that only a very light scalar field can contribute to the superhorizon CMB power spectrum. The contribution diminishes due to the decay of superhorizon mode functions, as given by \eq{ae}, which contrasts with the case of  $\m$-distortion, where high-energy modes also contribute, leading to a more significant effect. From a particle phenomenology perspective, the scalar field mass $m$ must be larger than the energy scales accessible to current accelerators. On the other hand, although the Higgs field ${\cal H}$ does not directly couple to the electromagnetic field in the Standard Model, an effective Higgs-photon coupling of the form ${\cal H}\,F^{\m\n}F_{\m\n}$ arises at one loop. The structure of this coupling differs from that in \eq{1}, and it would be interesting to investigate the impact of the Higgs field on photon propagation in the context of the present work.

The calculations in this paper rely on several core theoretical assumptions to evaluate the possible effects of vacuum polarization on the CMB. The photons are assumed to propagate along trajectories within the geometric-optics approximation. Since we analyze photon propagation from the last-scattering surface to the present day, this is an excellent approximation, as in the standard treatment of the Sachs-Wolfe effect. The charged scalar field is assumed to be light compared with the scale of inflation and to be initially in the Bunch-Davies vacuum. Consequently, superhorizon fluctuations of the scalar field are generated during inflation with a scale-free spectrum. The evolution of these scale-free perturbations after inflation is determined by the mode equation \eq{me}, in which the relative magnitudes of the scalar field mass $m$ and the Hubble parameter $H$ become crucial. The effect of superhorizon scalar-field fluctuations on a photon trajectory, mediated by the photon-scalar coupling, is assumed to be small, implying only a slight deviation from the null geodesic equation. This deviation is determined in the standard way, analogous to the derivation of the null geodesic equation within the geometric-optics approximation. We see that when an observable receives a correction that depends on the coincident limit of a two-point function, as in the case of the $\m$-distortion considered here, the final result is particularly sensitive to the choice of regularization prescription.

Clearly, the vacuum polarization effects are expected to be more significant at earlier times. In this case, even when photons couple to the thermal plasma, their frequencies are continuously shifted by the last term in \eq{w} as they freely travel between collisions. Therefore, when the mean free path is not very short, vacuum polarization has a chance to persistently influence the plasma, altering its mean energy. Investigating the consequences of this effect would be of interest as well.

\end{document}